\documentclass[12pt,a4paper]{article}
\usepackage{amssymb}
\usepackage{amsmath}
\usepackage{amsfonts}
\usepackage[pdftex]{graphicx}
\graphicspath{{./graphics/}}
\DeclareGraphicsExtensions{.pdf,.png,.jpg}
\usepackage{amscd}
\usepackage{a4wide}
\usepackage{microtype}
\usepackage{bbm}
\usepackage{slashed}
\usepackage{fancyhdr}
\usepackage{amsthm}
\usepackage{mathrsfs}
\usepackage{hyperref}
\usepackage{color} 
\usepackage{pstricks}
\usepackage{enumerate}
\usepackage{epstopdf}
\usepackage{venturis2}

\definecolor{darkred}{rgb}{0.8,0.1,0.1}
\hypersetup{
     colorlinks=true,         
     linkcolor=darkred,
     citecolor=blue,
}

\usepackage{comment}
\usepackage[all,cmtip]{xy}

\usepackage{marvosym}

\usepackage{geometry}
\newtheorem{theorem}{\rmfamily\bfseries{Theorem}}[section]

\theoremstyle{remark}

\newtheorem{remark}{Remark}

\def\un{1\kern-3pt \rm I}

\numberwithin{equation}{section}

\def\1{\mathbbm{1}}

\newcommand{\oR}{{\mathbb R}}

\newcommand{\oC}{{\mathbb C}}

\title{\bf{On the absence of bound states for a planar massless Brown-Ravenhall-type operator}}

\author{Magno B. Alves,$^{*}$ Oswaldo M. Del Cima$^\dag$ and Daniel H.T. Franco$^{\dag \ddag}$ \vspace{4mm}\\
       $^{*}$ Universidade Federal de Juiz de Fora, Departamento de Matem\'atica,\\
      Campus Universit\'ario, Bairro Martelos,\\
      Juiz de Fora, MG, Brasil, CEP: 36.036-900.\vspace{4mm}\\
      $^\dag$ Grupo de F\'\i sica-Matem\'atica e Teoria Qu\^antica dos Campos,\\
      Universidade Federal de Vi\c cosa, Departamento de F\'\i sica,\\
      Av. Peter Henry Rolfs s/n, Campus Universit\'ario,\\
      Vi\c cosa, MG, Brasil, CEP: 36570-900.\vspace{4mm}\\
      $^\ddag$ Centro de Estudos de F\'\i sica Te\'orica,\\
      Setor de F\'\i sica-Matem\'atica,\\
      Rua Rio Grande do Norte 1053/302, Funcion\'arios,\\
      Belo Horizonte, Minas Gerais, Brazil, CEP 30130-131.\vspace{4mm}\\
{\small e-mail: \texttt{magno\underline{~}branco@yahoo.com.br, oswaldo.delcima@ufv.br, daniel.franco@ufv.br}}
}

\date{\today}

\begin{document}

\maketitle

\begin{abstract}
We address the question of the existence of bound states for a suitably projected two-dimensional massless
Dirac operator in the presence of a Bessel-Macdonald potential (also known as $K_0$-potential potential),
raised in Ref.~\cite{WOE}. Based on Relativistic Hardy Inequality, we prove that this operator has no bound
states if $\gamma \leqslant \gamma_{\rm crit}$ (subcritical region), where $\gamma$ is a coupling constant.
\end{abstract}



\section{Introduction}
\label{Sec1}
\hspace*{\parindent}
In this short communication, we focus on the planar massless Dirac operator with Bessel-Macdonald potential
restricted to its positive spectral subspace (a Brown-Ravenhall-type operator); namely, in order to describe a
quasi-particle in the field of another quasi-particle and subject to relativistic effects, the suitably projected planar
massless Dirac operator with a Bessel-Macdonald potential is
\begin{align}
\boldsymbol{\cal B}(\boldsymbol{x})
=\Lambda_+\bigl(\boldsymbol{\cal D}_0(\boldsymbol{x})-\gamma K_0(\mu |\boldsymbol{x}|)\bigr)\Lambda_+\,\,,
\label{OpBR}
\end{align}
where $\boldsymbol{\cal D}_0$ is the free massless Dirac operator in $d=2$, $\gamma > 0$ is the coupling parameter
taken to be contained in the non-negative semi-axis $[0,\infty)$ and $K_0(\mu |\boldsymbol{x}|)$ is the Bessel-Macdonald
potential induced by
\begin{align*}
K_0(\mu |\boldsymbol{x}|)
=\frac {1}{2} \int _{0}^{\infty} e^{-{\frac {\pi \mu^2 |\boldsymbol{x}|^{2}}{\eta}}}
e^{-\frac{\eta}{4\pi}}~\frac{1}{\eta}\,d\eta\,\,.
\end{align*}
Here, $\mu > 0$ is a real parameter. Specifically, we shall address the question of the existence
of bound states for the operator (\ref{OpBR}) raised in Ref.~\cite{WOE}.

\begin{remark}[{\bf The origin of the operator (\ref{OpBR})}]
At this point, we recall that the physics of graphene has acted as a bridge between quantum field theory and
condensed matter physics due to the special quality of graphene quasi-particles behaving as massless two-dimensional
Dirac fermions~\cite{VKG}. In our case, the operator (\ref{OpBR}) emerges from a parity-preserving $U(1) \times U(1)$
massless ${\rm QED}_3$ proposed in~\cite{WOE} as a pristine graphene-like planar quantum electrodynamics model,
exhibiting two massive gauge bosons. In particular, the $K_0$-potential arises when analyzing the $s$- and $p$-wave M{\o}ller
(electron-polaron--electron-polaron) scattering amplitudes. While the $p$-wave state fermion-fermion (or antifermion-antifermion)
scattering potential shows to be repulsive whatever the values of the electric $(e)$ and chiral $(g)$ charges, for $s$-wave
scattering of fermion-fermion (or antifermion-antifermion), the interaction potential might be attractive provided $g^2 > e^2$
(see in Ref.~\cite{WOE} the details). Thus, the question of whether or not the attractive $s$-wave state potential favors
$s$-wave massless bipolarons (two-fermion bound states) shall be answered here by investigating into details the operator
(\ref{OpBR}), together with necessary and sufficient conditions which guarantee relativistic two-particle massless bound
states.
\end{remark} 

\section{Positive spectral subspace and reduction of spinors}
\label{Sec2}
\hspace*{\parindent}
Any attempt to describe bound states of spin-$1/2$ relativistic quasi-particles via the potential $K_0$
must take into account the problem of continuum dissolution \cite{sucher1}--\cite{sucher3}. A solution to
this problem is, according to Brown-Ravenhall~\cite{sucher1}, to consider the suitably projected planar
massless Dirac operator in the presence of a $K_0$-potential given by
\begin{align*}
\boldsymbol{\cal B}(\boldsymbol{x})
=\Lambda_+\bigl(\boldsymbol{\cal D}_0(\boldsymbol{x})-\gamma K_0(\mu |\boldsymbol{x}|)\bigr)\Lambda_+\,\,.
\end{align*}
Here, the operator $\boldsymbol{\cal D}_0$ is a first-order operator acting
on {\em spinor}-valued functions $\Psi(\boldsymbol{x})=(\psi_1(\boldsymbol{x}),\psi_2(\boldsymbol{x}))$, with $2$
components, of the space variable $\boldsymbol{x}=(x_1,x_2)$. We denote by $\oC^2$ the $2$-dimensional complex
vector space in which the values of $\Psi(\boldsymbol{x})$ lie. $\boldsymbol{\cal D}_0$ has the form
\begin{align*}
\boldsymbol{\cal D}_0=-i \boldsymbol{\sigma} \cdot \boldsymbol{\nabla}\,\,,
\end{align*}
where $\boldsymbol{\sigma}=(\sigma_1;\sigma_2)$ are the Pauli $2 \times 2$-matrices
\begin{align*}
\sigma_1=
\begin{pmatrix}
0 & 1 \\[3mm]
1 & 0
\end{pmatrix}\,\,,\quad
\sigma_2=
\begin{pmatrix}
0 & -i \\[3mm]
i & 0
\end{pmatrix}\,\,.
\end{align*}

$\Lambda_+ \overset{\rm def.}{=} \chi_{(0,\infty)}\bigl(\boldsymbol{\cal D}_0\bigr)$, where $\chi_{(0,\infty)}$
is the characteristic function of the interval $(0,+\infty)$, denotes the projection of $L_2(\oR^2;\oC^2)$ onto the
positive spectral subspace of $\boldsymbol{\cal D}_0$; namely,
\begin{align*}
\Lambda_+=\frac{1}{2}
\left({\un}_{2 \times 2}
+\frac{-i \boldsymbol{\sigma} \cdot \boldsymbol{\nabla}}
{\sqrt{-\Delta}}\right)\,\,,
\end{align*}
where $\Delta$ is the laplacian operator on $\oR^2$. Note that 
$\boldsymbol{\cal D}_0 \Lambda_+=\Lambda_+ \boldsymbol{\cal D}_0=\sqrt{-\Delta}\;\Lambda_+$.
The last equality is a consequence of the following fact: in Fourier variables, the projector $\Lambda_+$
is a multiplication operator given by the following expression:
\begin{align*}
\widehat{\Lambda}_+
=\frac{1}{2}
\left({\un}_{2 \times 2}
+\frac{\boldsymbol{\sigma} \cdot \boldsymbol{p}}
{|\boldsymbol{p}|}\right)\,\,.
\end{align*}

Hence, the Brown-Ravenhall-type operator is given formally as
\begin{align*}
\boldsymbol{\cal B}(\boldsymbol{x})
=\Lambda_+ \sqrt{-\Delta}\;\Lambda_+
-\gamma\;\Lambda_+ K_0(\mu |\boldsymbol{x}|) \Lambda_+\,\,,
\end{align*}
acting in $L_2(\oR^2;\oC^2)$, or, equivalently
\begin{align*}
\boldsymbol{\cal B}(\boldsymbol{x})
=\Lambda_+ \sqrt{-\Delta}-\gamma\;\Lambda_+ K_0(\mu |\boldsymbol{x}|)\,\,,
\end{align*}
acting in ${\mathscr H}_+ \overset{\rm def.}{=} \Lambda_+\bigl(L_2(\oR^2;\oC^2)\bigr)$.

The first step to prove the existence or absence of bound states for the operator (\ref{OpBR}) is a reduction of spinors.
We will follow the same strategy as Zelati-Nolasco~\cite{ZeNo}: we use the {\em Foldy-Wouthuysen transformation} (FW),
given by a unitary transformation $U_{\rm FW}$ which transforms the free massless Dirac operator into the diagonal form
(see details in~\cite{Binegar} for the case in $d=1+2$)
\begin{align*}
\boldsymbol{\cal D}_{\rm FW}=U_{\rm FW} \boldsymbol{\cal D}_0 U_{\rm FW}^{-1}
=\begin{pmatrix}
\sqrt{-\Delta} & 0 \\[3mm]
0 & -\sqrt{-\Delta}
\end{pmatrix}
=\sigma_3 \sqrt{-\Delta}\,\,,
\end{align*}
where $\sigma_3$ is the Pauli $2 \times 2$-matrix
\begin{align*}
\sigma_3=
\begin{pmatrix}
1 & 0 \\[3mm]
0 & -1
\end{pmatrix}\,\,.
\end{align*}

\begin{remark}
With the usual quantization rule $\boldsymbol{p} \mapsto -i \boldsymbol{\nabla}$ (here the units are chosen so
that $\hbar=c=1$), let us recall that to the operator $\sqrt{-\Delta}$ can be defined for all $\Psi \in H^1(\oR^2;\oC^2)$
as the inverse Fourier transform of the $L_2$-function $|\boldsymbol{p}|\widehat{\Psi}(\boldsymbol{p})$ (where $\widehat{\Psi}$
denotes the Fourier transform of $\Psi$). To $\sqrt{-\Delta}$ we can associate the following quadratic form
\begin{align*}
q_{_H}(\Phi,\Psi)
\overset{\rm def.}{=} \langle \Phi,\sqrt{-\Delta}\,\Psi \rangle=\frac{1}{(2\pi \hbar)^{2}} \int_{\oR^2}
|\boldsymbol{p}|~\overline{\widehat{\Phi}}(\boldsymbol{p}) \widehat{\Psi}(\boldsymbol{p})~d\boldsymbol{p}\,\,,
\end{align*}
which can be extended to all functions $\Phi,\Psi$ in the form domain $\boldsymbol{\mathfrak{Q}}(\sqrt{-\Delta})=H^{1/2}(\oR^2;\oC^2)$,
where
\begin{align*}
H^{1/2}(\oR^2;\oC^2)=\left\{\Psi \in L_2(\oR^2;\oC^2) \mid 
\int_{\oR^2} (1+|\boldsymbol{p}|^2)^{1/2} |\widehat{\Psi}(\boldsymbol{p})|^2~d\boldsymbol{p} < \infty\right\}\,\,.
\end{align*}
\label{Note01}
\end{remark}

Under the FW-transformation the projector $\Lambda_+$ becomes simply
\begin{align*}
\Lambda_{+{\rm FW}} \overset{\rm def.}{=} U_{\rm FW} \Lambda_+ U_{\rm FW}^{-1}
=\frac{1}{2}
\left({\un}_{2 \times 2}+\sigma_3\right)\,\,.
\end{align*}
Therefore the positive energy subspace for $\boldsymbol{\cal D}_{\rm FW}$ is simply given by
\begin{align*}
{\mathscr H}_+=\left\{\Psi=\binom{\psi}{0} \in L_2(\oR^2;\oC^2) \mid \psi \in L_2(\oR^2;\oC) \right\}
\end{align*}

In the FW-representation the associated quadratic form acting on ${\mathscr H}_+$ is defined by
\begin{align}
\langle \varphi,\boldsymbol{\cal B}_{\rm FW} \psi \rangle_{L_2(\oR^2;\oC)}=
\left\langle \varphi,\sqrt{-\Delta}\,\psi \right\rangle_{L_2(\oR^2;\oC)}
+ \langle \varphi,V_{\rm FW} \psi \rangle_{L_2(\oR^2;\oC)}\,\,,
\label{OpBRFW}
\end{align}
for any $\varphi,\psi \in H^{1/2}(\oR^2;\oC)$, where $V_{\rm FW} \psi=Q^*U_{\rm FW}VU_{\rm FW}^{-1}Q\psi$, with
$V(\boldsymbol{x})=-\gamma K_0(\beta |\boldsymbol{x}|)$ and
\begin{align*}
&Q:  \oC \to \oC^2\,\,,\quad Q(z_1)=(z_1,0)\,\,, \\[3mm]
&Q^*:\oC^2 \to \oC\,\,,\quad Q^*(z_1,z_2)=z_1\,\,,
\end{align*}
so that
\begin{align*}
\left\langle \varphi,\sqrt{-\Delta}\,\psi \right\rangle_{L_2(\oR^2;\oC)}
&=\left\langle \Lambda_+U_{\rm FW}^{-1}Q\varphi,\boldsymbol{\cal D}_0 \Lambda_+U_{\rm FW}^{-1}Q\psi \right\rangle_{L_2(\oR^2;\oC^2)} \\[3mm]
&=\left\langle \Lambda_+U_{\rm FW}^{-1} \binom{\varphi}{0},\boldsymbol{\cal D}_0 \Lambda_+U_{\rm FW}^{-1} \binom{\psi}{0}\right\rangle_{L_2(\oR^2;\oC^2)}\,\,,
\end{align*}
and
\begin{align*}
\left\langle \varphi,V_{\rm FW}\psi \right\rangle_{L_2(\oR^2;\oC)}
&=\left\langle U_{\rm FW}^{-1}Q\varphi,V U_{\rm FW}^{-1}Q\psi \right\rangle_{L_2(\oR^2;\oC^2)} \\[3mm]
&=\left\langle \Lambda_+U_{\rm FW}^{-1} \binom{\varphi}{0},V\Lambda_+U_{\rm FW}^{-1} \binom{\psi}{0}\right\rangle_{L_2(\oR^2;\oC^2)}\,\,.
\end{align*}
Note that $U_{\rm FW}^{-1}Q\varphi=\Lambda_+U_{\rm FW}^{-1}Q\varphi \in \Lambda_+\bigl(L_2(\oR^2);\oC^2\bigr)$
for any $\varphi \in L_2(\oR^2;\oC)$.

Thus, for any $\psi$ in the positive spectral subspace, the expectation of $\boldsymbol{\cal B}$ in the state $\Psi$
in the FW-representation is associated with the quadratic form
\begin{align}
\langle \psi,\boldsymbol{\cal B}_{\rm FW} \psi \rangle=
\left\langle \psi,\sqrt{-\Delta}\,\psi \right\rangle
-\gamma \langle \psi,K_0(\mu |\boldsymbol{x}|) \psi \rangle\,\,,
\label{OpBRFWa}
\end{align}
with form domain $\boldsymbol{\mathfrak{Q}}\bigl(\sqrt{-\Delta}\bigr)=H^{1/2}(\oR^2;\oC)$.
Hence, the transition from $\Psi \in L_2(\oR^2;\oC^2)$ to the reduced spinor $\psi \in L_2(\oR^2;\oC)$ through
the introduction of the operator $\boldsymbol{\cal B}_{\rm FW}$ is possible because we are working in
${\mathscr H}_+$. Naturally, the map $\langle \Psi,\boldsymbol{\cal B} \Psi \rangle \to \langle \psi,\boldsymbol{\cal B}_{\rm FW} \psi \rangle$,
where $\Psi \in L_2(\oR^2;\oC^2)$ and $\psi \in L_2(\oR^2;\oC)$, determines a unitary equivalence between the operators
$\boldsymbol{\cal B}$ and $\boldsymbol{\cal B}_{\rm FW}$. Hence, the absence of bound states for the operator $\boldsymbol{\cal B}_{\rm FW}$
implies  the absence of bound states for the operator $\boldsymbol{\cal B}$. So, from now on, we will work directly with the operator
\begin{align}
\boldsymbol{\cal B}_{\rm FW}=\sqrt{-\Delta}-\gamma K_0(\mu |\boldsymbol{x}|)\,\,.
\label{OMDBM}
\end{align}

\section{Absence of bound states}
\label{sec3}
A standard problem of spectral theory in quantum mechanics is to obtain conditions on a
potential in order to guarantee that this potential has bound states. In this section, our goal is to investigate
whether the potential of the Bessel-Macdonald type, $V(\boldsymbol{x})=\gamma K_0(\mu |\boldsymbol{x}|)$,
can lead to bound states of massless Dirac quasi-particles. Our investigation is based on the following
result~\cite[Lemma 8.2]{LS}:

\begin{theorem}[Relativistic Hardy Inequality]
Let $d \geqslant 2$, and let $\psi$ be a function
in $H^{1/2}(\oR^d)$. Then
\begin{align}
\int_{\oR^d} \frac{1}{|\boldsymbol{x}|}|\psi(\boldsymbol{x})|^2\,d\boldsymbol{x}
\leqslant {\cal C}_d^2 \int_{\oR^d} |\boldsymbol{p}| |\widehat{\psi}(\boldsymbol{p})|^2\,d\boldsymbol{p}
={\cal C}_d^2 \int_{\oR^d} \overline{\psi}(\boldsymbol{x}) \sqrt{-\Delta}\;\psi(\boldsymbol{x})\,d\boldsymbol{x}\,\,,
\label{LS1}
\end{align}
where the best possible value of the constant ${\cal C}_d$ is
\begin{align*}
{\cal C}_d=\frac{\Gamma\left(\frac{d-1}{4}\right)}{\sqrt{2}\,\Gamma\left(\frac{d+1}{4}\right)}\,\,.
\end{align*}
The equality is only attained if $\psi=0$, {i.e.}, for any bigger constant the inequality fails for some function
in $H^{1/2}(\oR^d)$.
\label{LS2}
\end{theorem}

\begin{remark}
This inequality goes back to Kato~\cite[Eq.(V.5.33)]{Kato} and Herbst~\cite[Theorem 2.5]{Herbst}. See
also~\cite[Theorem 1.7.1]{BaEvLe}.
\end{remark}

By applying Theorem \ref{LS2} to $\langle \psi,\boldsymbol{\cal B}_{\rm FW} \psi \rangle$ we obtain
\begin{align}
\langle \psi,\boldsymbol{\cal B}_{\rm FW} \psi \rangle
\geqslant \int_{\oR^2} \overline{\psi}(\boldsymbol{x})
\left(\frac{1}{{\cal C}_2^2} \frac{1}{|\boldsymbol{x}|}-\gamma K_0(\mu |\boldsymbol{x}|)\right)
\psi(\boldsymbol{x})\,d\boldsymbol{x}\,\,.
\label{compare0}
\end{align}

Next, we shall use the following rough estimate
\begin{align*}
K_0(\mu |\boldsymbol{x}|)
&=\frac {1}{2} \int _{0}^{\infty} e^{-{\frac {\pi \mu^2 |\boldsymbol{x}|^{2}}{\eta}}}
e^{-\frac{\eta}{4\pi}}~\frac{1}{\eta}\,d\eta \nonumber \\[3mm]
&=\frac {1}{2 |\boldsymbol{x}|^\alpha} \int _{0}^{\infty} \frac{|\boldsymbol{x}|^\alpha}{\eta^{\alpha/2}}
e^{-{\frac {\pi \mu^2 |\boldsymbol{x}|^{2}}{\eta}}}
e^{-\frac{\eta}{4\pi}}~\frac{1}{\eta^{1-\alpha/2}}\,d\eta \nonumber \\[3mm]
&\leqslant \frac {1}{2 |\boldsymbol{x}|^\alpha} 
\left(\sup_{t > 0} t^{\alpha/2} e^{-\pi \mu^2 t}\right)
\int _{0}^{\infty} e^{-\frac{\eta}{4\pi}}~\frac{1}{\eta^{1-\alpha/2}}\,d\eta\,\,.
\end{align*}
Now, since
\begin{enumerate}[(i)]
\item $\displaystyle{t^{\alpha/2} e^{-\pi \mu^2 t} > 0\,\,,\quad \forall\;t >0}$\,\,,
\item $\displaystyle{\lim_{t \to 0^{+}}\left(t^{\alpha/2} e^{-\pi \mu^2 t}\right)=0=\lim_{t \to +\infty}\left(t^{\alpha/2} e^{-\pi \mu^2 t}\right)}$\,\,,
\item $\displaystyle{\frac{d}{dt}\left(t^{\alpha/2} e^{-\pi \mu^2 t}\right)=0 \iff t=\frac{\alpha}{2 \pi \mu^2}}$\,\,,
\end{enumerate}
we conclude that 
\begin{align*}
\sup_{t > 0} t^{\alpha/2} e^{-\pi \mu^2 t}
=t^{\alpha/2} e^{-\pi \mu^2 t}\Bigl|_{t=\frac{\alpha}{2 \pi \mu^2}}
=\left(\frac{\alpha}{2\pi \mu^2}\right)^{\alpha/2} e^{-\alpha/2}\,\,.
\end{align*}
Furthermore, according the Table of Integrals of Gradshtein-Ryzhik~\cite[{\bf 3.381}, $4.$, p.346]{GR}, it follows that
\[
\int _{0}^{\infty} e^{-\frac{\eta}{4\pi}}~\frac{1}{\eta^{1-\alpha/2}}\,d\eta
=(4\pi)^{\alpha/2} \Gamma(\alpha/2)\,\,.
\]
Hence, we have
\begin{align*}
K_0(\mu |\boldsymbol{x}|) \leqslant \frac {C_{\alpha,\mu}}{|\boldsymbol{x}|^\alpha}\,\,,
\end{align*}
where
\begin{align*}
C_{\alpha,\mu}=\frac{1}{2} \left(\frac{\alpha}{2\pi \mu^2}\right)^{\alpha/2} 
(4\pi)^{\alpha/2} \Gamma(\alpha/2) e^{-\alpha/2}\,\,.
\end{align*}

If we take $\alpha=1$, then
\begin{align*}
K_0(\mu |\boldsymbol{x}|) \leqslant  \frac {C_{1,\mu}}{|\boldsymbol{x}|}\,\,.
\end{align*}
This implies, according to relativistic Hardy inequality, that we have for $d=2$
\begin{align}
\int_{\oR^2} \overline{\psi}(\boldsymbol{x})
\left(\frac{1}{{\cal C}_2^2} \frac{1}{|\boldsymbol{x}|}-\gamma K_0(\mu |\boldsymbol{x}|)\right)
\psi(\boldsymbol{x})\,d\boldsymbol{x}
\geqslant
\frac{1}{{\cal C}_2^2} \left(1-\gamma {\cal C}_2^2 C_{1,\mu} \right) 
\int_{\oR^2} \frac{1}{|\boldsymbol{x}|}
|\psi(\boldsymbol{x})|^2\,d\boldsymbol{x}\,\,.
\label{compare}
\end{align}
Note that
\begin{align*}
{\cal C}_2^2 C_{1,\mu}
=\frac{1}{\sqrt{2}} \Gamma(1/2) e^{-1/2} \left(\frac{\Gamma(1/4)}{\sqrt{2}\,\Gamma(3/4)}\right)^2 \frac{1}{\mu}\,\,.
\end{align*}

We can simplify the expression of this constant, taking into account the relationship that exists between the gamma function
and the beta function. Indeed, it follows that
\begin{align*}
{\cal C}_2^2 C_{1,\mu}
=\frac{[B(1/2,1/4)]^2}{\sqrt{8 \pi e}}\;\frac{1}{\mu}
=\frac{[\Gamma(1/4)]^4}{2 (2\pi)^{3/2} e^{1/2}}\;\frac{1}{\mu}\,\,.
\end{align*}
In the last equality, we use the well-known expression
\begin{align*}
B(x,y)=2 \int_0^{\pi/2} (\cos \varphi)^{2x-1} (\sin \varphi)^{2y-1}\;d\varphi\,\,, 
\end{align*}
and the Table of Integrals of Gradshtein-Ryzhik~\cite[{\bf 3.621}, $7.^*$, p.395]{GR} in order to calculate the value of the 
function $B(1/2,1/4)$.

At this point, we remark a number of interesting properties of planar {\em massive} Dirac operator with Bessel-Macdonald potential,
restricted to its positive spectral subspace, which have been obtained in Ref.~\cite{MOD}. For instance, the following have been
established.
\begin{enumerate}
\item $\boldsymbol{\cal B}_{\rm FW}$ is shown to be bounded below if, and only if,
$\gamma \leqslant \gamma_{\rm crit}=({\cal C}_2^2 C_{1,\mu})^{-1}$ (a property also referred to as stability of matter).
$\boldsymbol{\cal B}_{\rm FW}$ is, in fact, shown to be positive in~\cite[Proposition 4.5]{MOD}, the estimate
(in appropriate units) $\boldsymbol{\cal B}_{\rm FW} \geqslant m(1-\gamma \gamma_{\rm crit}^{-1})$ being obtained.

\item $\boldsymbol{\cal B}_{\rm FW}$ is self-adjoint on the form domain $H^{1/2}(\oR^2;\oC)$ if
$\gamma < \gamma_{\rm crit}$~\cite[Proposition 3.4]{MOD}. For the critical value $\gamma=\gamma_{\rm crit}$,
the Friedrichs Extension Theorem guarantees that the quadratic form $\langle \psi,\boldsymbol{\cal B}_{\rm FW} \psi \rangle$
is a closable quadratic form and its closure is the quadratic form of a unique self-adjoint operator associated with
$\boldsymbol{\cal B}_{\rm FW}$. Thereby, the critical coupling constant $\gamma_{\rm crit}$ can be mathematically thought
of as that coupling constant where a natural definition of self-adjointness ceases to exist.

\item The essential spectrum $\sigma_{\rm ess}(\boldsymbol{\cal B}_{\rm FW})$ is proved in~\cite[Theorem 4.1]{MOD}
to coincide with $[m,\infty)$ when $\gamma \leqslant \gamma_{\rm crit}$. Possible embedded eigenvalues in the essential
spectrum $\sigma_{\rm ess}(\boldsymbol{\cal B}_{\rm FW})$ are absent~\cite[Lemma 4.4]{MOD}. In particular, all eigenvalues
are non-negative, {\em i.e.}, in $[0,m)$ the discrete spectrum $\sigma_{\rm disc}(\boldsymbol{\cal B}_{\rm FW})$ consists
of an infinite number of isolated eigenvalues of finite multiplicity.
\end{enumerate}

Returning to the massless case, since the difference between the operator $\sqrt{-\Delta+m^2}$ and $\sqrt{-\Delta}$
is bounded, more precisely $-\Delta+m^2 \geqslant \sqrt{-\Delta+m^2} \geqslant \sqrt{-\Delta}$, the stability described in
Remark 1 is the same as the stability of operator (\ref{OMDBM}). Note that for $m=0$ the bound in Remark 1 shows
positivity directly. Moreover, according to Remark 2, the operator (\ref{OMDBM}) is self-adjoint on the form domain
$H^{1/2}(\oR^2;\oC)$ if $\gamma < \gamma_{\rm crit}$. Finally, with respect to Remark 3, the operator (\ref{OMDBM})
has the essential spectrum to coincide with $[0,\infty)$ when $\gamma \leqslant \gamma_{\rm crit}$. In this case,
all eigenvalues, $\lambda$, should be {\em negative}. But, according to (\ref{compare}), this would only be possible if
$\gamma > \gamma_{\rm crit}$ (supercritical region). We have nothing to say about this, since this would imply the
non-self-adjointness of the operator $\boldsymbol{\cal B}_{\rm FW}$ and, therefore, the absence of dynamics by the
celebrated Stone's Theorem~\cite[Theorem VIII.8]{RS}. This leaves the possibility of $\lambda=0$. Then, suppose that
0 is an eigenvalue of $\boldsymbol{\cal B}_{\rm FW}$ with corresponding eigenfunction $\psi$. Thus, the right-hand
side of (\ref{compare0}) must be zero. But this would imply that there is equality in (\ref{LS1}) with 
$\psi \not= 0$, which is not possible.

In short, bound states for the operator (\ref{OMDBM}) should occur whenever a quasi-particle in the field of another
quasi-particle cannot move to infinity. That is, the quasi-particle should be confined or bound at all energies to move within a finite
and limited region of space. The operator (\ref{OMDBM}) in this region admits only solutions with eigenvalues that are in
the $\sigma_{\rm disc}(\boldsymbol{\cal B}_{\rm FW})$, which in this case is empty for $\gamma \leqslant \gamma_{\rm crit}$.
Hence, for the planar massless Dirac operator with Bessel-Macdonald potential, restricted to its positive spectral subspace,
there are no bound states in the subcritical region. Thus, with the analysis carried out above, we are in a position to answer
the question raised in  Ref.~\cite{WOE} by means of the following

\begin{theorem}
For $\gamma \leqslant \gamma_{\rm crit}$, the massless planar Dirac operator with Bessel-Macdonald potential, 
restricted to its positive spectral subspace, has no eigenvalues and therefore has no bound states.
\end{theorem}

\section*{Author's Contributions}
\hspace*{\parindent}
All authors contributed equally to this work. On behalf of all authors, the corresponding author states that
there is no conflict of interest. 

\section*{Data Availability}
\hspace*{\parindent}
The data that support the findings of this study are available from the corresponding author upon reasonable request.



\end{document}